
\magnification = 1200
\def\lapp{\hbox{$ {
\lower.40ex\hbox{$<$}
\atop \raise.20ex\hbox{$\sim$}
}
$}  }
\def\rapp{\hbox{$ {
\lower.40ex\hbox{$>$}
\atop \raise.20ex\hbox{$\sim$}
}
$}  }
\def\barre#1{{\not\mathrel #1}}
\def\krig#1{\vbox{\ialign{\hfil##\hfil\crcr
$\raise0.3pt\hbox{$\scriptstyle \circ$}$\crcr\noalign
{\kern-0.02pt\nointerlineskip}
$\displaystyle{#1}$\crcr}}}
\def\upar#1{\vbox{\ialign{\hfil##\hfil\crcr
$\raise0.3pt\hbox{$\scriptstyle \leftrightarrow$}$\crcr\noalign
{\kern-0.02pt\nointerlineskip}
$\displaystyle{#1}$\crcr}}}
\def\ular#1{\vbox{\ialign{\hfil##\hfil\crcr
$\raise0.3pt\hbox{$\scriptstyle \leftarrow$}$\crcr\noalign
{\kern-0.02pt\nointerlineskip}
$\displaystyle{#1}$\crcr}}}

\def\Tr{\,{\rm Tr }\,}

\def\g5{\gamma_5}

\def\lp1{{\cal L}_{\pi N}^{(1)}}
\def\lp2{{\cal L}_{\pi N}^{(2)}}
\def\lp3{{\cal L}_{\pi N}^{(3)}}

\topskip=0.60truein
\leftskip=0.18truein
\vsize=8.8truein
\hsize=6.5truein
\tolerance 10000
\hfuzz=20pt

\baselineskip 14pt plus 1pt minus 1pt
\pageno=0
\centerline{\bf LOW--ENERGY THEOREMS FOR WEAK PION PRODUCTION}
\vskip 48pt
\centerline{V. Bernard$^1$, N. Kaiser$^2$ and
Ulf-G. Mei{\ss}ner$^1$}
\vskip 16pt
\centerline{$^1${\it Centre de Recherches Nucl\'{e}aires et Universit\'{e}
Louis Pasteur de Strasbourg}}
\centerline{\it Physique Th\'{e}orique,
BP 20Cr, 67037 Strasbourg Cedex 2, France}
\vskip  4pt
\centerline{$^2${\it Physik Department T30,
Technische Universit\"at M\"unchen}}
\centerline{\it
James
Franck Stra{\ss}e,
D-85747 Garching, Germany}
\vskip  4pt
\vskip 12pt
\vskip 1.0in
\centerline{\bf ABSTRACT}
\medskip
\noindent We derive novel low--energy theorems for single pion production off
nucleons through the isovector axial current. We find that the $k^2$-dependence
of the multipole $L_{0+}^{(+)}$ at threshold is given by the nucleon scalar
form factor, namely $\sigma(k^2-M_\pi^2 ) /(3 \pi  M_\pi F_\pi )$. The
relation to PCAC results for soft pions including electroweak form factors is
also clarified.
\medskip
\vfill
\noindent CRN--93/56  \hfill December 1993
\vskip 12pt
\eject
\baselineskip 14pt plus 1pt minus 1pt
\noindent 1. Single pion production off nucleons by real or virtual photons
gives important information about the structure of the nucleon. As stressed
in particular by Adler [1], weak pion production involves the isovector axial
amplitudes and      a unified treatment of pion photo-, electro- and weak
production allows to relate information from electron--nucleon and
neutrino--nucleon scattering experiments. In this spirit, we will consider
here  pion production through the isovector axial current in the
threshold region. This is a straightforward generalization of the results for
photo- and electroproduction given in refs.[2,3,4,5,6]. This goes beyond the
work of Adler [1], Adler and Dothan [7] and Nambu, Luri{\'e} and Shrauner [8],
who have considered soft pion emission induced by weak interactions making
use of PCAC and gauge invariance, relating            certain electroweak
form factors of the nucleon to particular threshold multipole amplitudes.
The methods used by these authors are, however,
             only exact in the chiral limit of vanishing pion mass and a
systematic evaluation of the explicit chiral symmetry breaking through the
quark (pion) masses has not yet been considered. To fill in this gap, we will
make use of chiral perturbation theory [9,10,11] which allows to systematically
explore the consequences of the spontaneous and explicit chiral symmetry
breaking in QCD by evaluating a finite number of Feynman diagrams.
To include baryons (like nucleons) in the effective field
theory, it is advantegous to consider the extreme non--relativistic limit for
the baryons as pointed out by Jenkins and Manohar [12] and discussed in detail
for the two-flavour case in ref.[13].
The purpose of the present note is to give novel relations between
various axial threshold multipole amplitudes and physical observables
like electroweak form factors, S--wave pion--nucleon scattering lenghts and,
in particular, the nucleon scalar form factor, $\sigma(t) \sim\,<N|\hat m (\bar
u u + \bar d d)|N>$.
\bigskip \noindent
2. We consider processes of the type $\nu(k_1) + N(p_1)
\to l(k_2) + N(p_2) + \pi^a(q)$, which involve the isovector vector
and axial--vector currents. Here, we will focus on the pion production
induced by the axial current, $A_\mu^b = \bar q \gamma_\mu \gamma_5 (
\tau^b / 2) q$ in terms of the $u$ and $d$ quark fields. Denoting by
$k = k_1 - k_2$ the four--momentum of the axial current, the pertinent
Mandelstam variables are $s=(p_1+k)^2$, $t=(q-k)^2$ and $u=(p_1 -q)^2$
subject to the constraint $s+t+u = 2m^2 + M_\pi^2 +k^2$ ($m$ and
$M_\pi$ denote the nucleon and the pion mass, in order), see fig.1.
The pertinent matrix element decomposes into an isospin--even and
an isospin--odd part (analogous to the $\pi N$ scattering amplitude),
$${_{\rm out}<} N(p_2), \pi^a(q) | A^b \cdot \epsilon | N(p_1)>_{\rm in}
\, = \, i \delta^{ab}\, T^{(+)} \cdot \epsilon \, \,- \, \, \epsilon^{abc} \,
 \tau^c \,
T^{(-)} \cdot \epsilon    \eqno(1)$$
where $\epsilon_\mu$ is the axial polarization vector, $\epsilon_\mu
\sim \bar{u}_l \gamma_\mu \gamma_5 u_\nu$. Notice that one can use the
Dirac equation to transform terms of the type $\epsilon \cdot k$ into
lepton mass terms via
$$\epsilon \cdot k \sim \bar{u}_l \barre{k} \gamma_5 u_\nu = - m_l
\bar{u}_l  \gamma_5 u_\nu \quad . \eqno(2)$$
This means that in the approximation of zero lepton mass, one has
$\epsilon \cdot k = 0$ and all diagrams where the axial source  couples
directly to a pion line vanish. The general Dirac structure for the transition
current involves the eight operators ${\cal O}_1 = (\barre \epsilon\,
\barre q - \barre q\,\barre \epsilon)/2 , \, {\cal O}_2 = \epsilon \cdot q,\,
{\cal O}_3 = \barre \epsilon,\, {\cal O}_4 = \epsilon \cdot (p_1 + p_2)/2,\,
{\cal O}_5 = \barre k\,\epsilon \cdot q,\, {\cal O}_6 = \barre k\,\epsilon\cdot
(p_1 + p_2) /2 ,\, {\cal O}_7 = \epsilon \cdot k,\, {\cal O}_8 = \barre k\,
\epsilon \cdot k$ which are accompanied by invariant functions denoted
$A_i^{(\pm )} (s,u)$ $(i=1, \ldots ,8)$ [1]. At threshold in the $\pi N$ center
of mass frame, one can express the
pertinent matrix element in terms of six S--wave multipoles, called
$L_{0+}^{(\pm )}$,
$M_{0+}^{(\pm )}$ and
$H_{0+}^{(\pm )}$,
$$T^{(\pm)}\cdot \epsilon = \bar{u}_2 \sum_{i=1}^8 {\cal O}_i A_i^{(\pm)}
(s_{\rm th}, u_{\rm th}) u_1 = 4 \pi ( 1 + \mu ) \chi_2^\dagger \bigl[
\epsilon_0 L_{0+}^{(\pm )} + \epsilon \cdot k H_{0+}^{(\pm )} + i \vec{\sigma}
\cdot (\hat{k} \times \vec{\epsilon} ) M_{0+}^{(\pm )} \bigr] \chi_1 \eqno(3)$$
with $\chi_{1,2}$ two--component Pauli  spinors and $\mu = M_\pi / m$
the ratio of the pion and nucleon mass.  The calculation of
$L_{0+}^{(\pm )}$ and  $M_{0+}^{(\pm )}$ can most easily be done by setting
the lepton mass (i.e. $\epsilon \cdot k$) to zero. The way $L_{0+}^{(\pm )}$
is defined in (3) avoids a kinematical singularity at $k^2 = -M_\pi(2m+M_\pi)$.
At threshold, one can express
$L_{0+}$, $M_{0+}$ and $H_{0+}$ (suppressing isospin indices) through
the invariant amplitudes $A_i (s,u)$ via
$$\eqalign{
M_{0+} &= {\sqrt{\mu^2 - \nu} \over 8 \pi (1+ \mu )^{3/2}}\bigl\lbrace
m \mu A_1 - A_3 \bigr\rbrace \cr
L_{0+} &= {\sqrt{(2+\mu)^2 - \nu} \over
8 \pi (1+ \mu )^{3/2}} \biggl\lbrace -{\mu ( 2 + \mu ) + \nu \over
(2+\mu)^2 - \nu} \mu m A_1 + m \mu A_2 \cr
&\quad + {2(1+\mu)(2+\mu) \over (2+\mu)^2 - \nu} A_3 + m\bigl(1+ {\mu \over 2}
\bigr)A_4 + \mu^2 m^2 A_5 + m^2 \mu \bigl( 1+ {\mu \over 2} \bigr) A_6
\biggr\rbrace \cr
H_{0+} &= {\sqrt{(2+\mu)^2 - \nu} \over
8 \pi (1+ \mu )^{3/2}} \biggl\lbrace {2 ( 1 + \mu )  \over
(2+\mu)^2 - \nu} \bigl(\mu  A_1 - {1 \over m} A_3 \bigr) - {1 \over 2} A_4
- {m \over 2} \mu A_6 + A_7 + \mu m A_8
\biggr\rbrace
 \cr} \eqno(4)$$
where the $A_i (s,u)$ are evaluated at threshold $s_{\rm th} = m^2
(1+ \mu )^2$ and $u_{\rm th} = m^2 [1+ \mu (\nu - 1 - \mu )] /(1 +
\mu )$ and $\nu = k^2 / m^2$. We seek an expansion of these
threshold multipoles in powers of $\mu$ and $\nu$ up to and including
order ${\cal O}(\mu^2 ,\nu )$ (modulo logarithms).
\bigskip
\noindent 3. To work out the chiral expansion of the various threshold
multipoles, we invoke CHPT, where the fundamental Lagrangian ${\cal L}_{\rm
QCD}$ is substituted by an effective theory formulated in terms of the
aymptotically observed fields. We consider the two--flavor case in the isospin
limit.
The Goldstone pion fields are collected in the SU(2) matrix
$$U(x) = \sqrt{1 - \vec{\pi}^2 (x)/F^2} +
i{\vec \tau}\cdot{\vec \pi}(x) / F
= u^2 (x) \eqno(5)$$
with $F$ the pion decay constant in the chiral limit. The
nucleons are considered as very heavy, i.e. non--relativistically [12,13].
This allows for a projection onto velocity eigenstates and one can eliminate
the troublesome nucleon mass term from the effective theory. Thereby a string
of vertices of increasing power in $1/m$ is generated, with $m$ the nucleon
mass (in the chiral limit).\footnote{$^{1)}$}{Later on, we will identify $m$
with the physical nucleon mass.  Similarly, all quantities in the lowest order
effective theory differ by terms of the order $\hat m = (m_u + m_d) / 2$
from their physical values. Our final results will be expressed
entirely in terms of physical quantities. This is a consistent
procedure to the order we are working.}
The effective Lagrangian to order ${\cal
O}(q^3)$, where $q$ denotes a genuine small momentum or a meson
(quark) mass, reads
$$\eqalign{{\cal L}_{\pi N} &= {\cal L}_{\pi N}^{(1)} +
{\cal L}_{\pi N}^{(2)} + {\cal L}_{\pi N}^{(3)} \cr
{\cal L}_{\pi N}^{(1)} &= {\bar H} ( i v\cdot D + g_A S \cdot u ) H \cr
{\cal L}_{\pi N}^{(2)} &= {\bar H} \biggl\lbrace  -{1\over 2m} D \cdot D +
{1 \over 2m} (v \cdot D)^2 +   c_1   \Tr \chi_+   + \bigl( c_2 -{g_A^2
\over 8 m} \bigr)  v\cdot u\, v\cdot u \cr & \qquad \quad + c_3\,  u\cdot u -
{i g_A \over 2m} \lbrace S \cdot D , v \cdot u \rbrace - { i  \over 4 m}(1 +
\kappa_V) [ S^\mu , S^\nu ] f^+_{\mu \nu}\biggr\rbrace H \cr} \eqno(6)$$
with
$$u_\mu = i u^\dagger \nabla_\mu U u^\dagger \eqno(7)$$
where $H$ denotes the heavy nucleon field,
$v_\mu$ the four--velocity of $H$, $S_\mu$ the covariant spin--operator
subject to the constraint $v \cdot S = 0$, $\nabla_\mu$ the covariant
derivative acting on the pions, $D_\mu = \partial_\mu + \Gamma_\mu$ the chiral
covariant derivative for nucleons and we adhere to the notations of
refs.[13,14]. The superscripts (1,2,3) denote the chiral power. The lowest
order effective Lagrangian is of order ${\cal O}(q)$. The one loop contribution
is suppressed with respect to the tree level by $q^2$ thus contributing at
${\cal O}(q^3)$. Finally, there are contact terms of order $q^2$ with
coefficients not fixed by chiral symmetry.
The low--energy constants $c_1$ and $c_2 + c_3$ can be
expressed in terms of the $\pi N$ $\sigma$--term $\sigma(0)$ and the isospin
even  S--wave    $\pi N$ scattering length $a^+$, see eqs.(9,10) in ref.[15].
For the case at hand, ${\cal L}_{\pi N}^{(3)}$ only leads to
$k^2$--independent contributions to the threshold multipoles
(apart from kinematical corrections) and
we therefore do not explicitely give all these terms here. We also have not
exhibited the standard meson Lagrangian ${\cal L}_{\pi \pi}^{(2)}$, the
non--linear $\sigma$--model coupled to external (axial and pseudoscalar)
sources.
\bigskip
\noindent 4. The calculation of the tree level amplitudes from ${\cal
L}^{(1)}_{\pi N}$ and ${\cal L}^{(2)}_{\pi N}$ is straightforward
and the corresponding threshold multipoles can entirely be expressed in terms
of $F_\pi$, the axial--vector coupling $g_A$, the isovector anomalous magnetic
moment $\kappa_V$ and the physical quantities $a^+,\, a^-,\, \sigma(0)$
calculated to this order. One finds
$$\eqalign{
M_{0+}^{(+)}
 &= {\sqrt{M_\pi^2 - k^2} \over 16 \pi m F_\pi} \,g_A^2
\, , \qquad \cr
M_{0+}^{(-)}
&= {\sqrt{M_\pi^2 - k^2} \over 16 \pi m F_\pi} (1 +\kappa_V - g_A^2 )\,, \cr
L_{0+}^{(+)} &=-{M_\pi \over 2 \pi F_\pi}(c_2 + c_3) =
{\sigma(0) \over 4 \pi M_\pi F_\pi} - {a^+ F_\pi \over M_\pi} - {g_A^2 M_\pi
\over 16 \pi m F_\pi}   , \,\, \cr
L_{0+}^{(-)}&= { 1 \over 8 \pi F_\pi} \biggl[ -1 + {M_\pi \over 2m} (g_A^2 + 1)
\biggr] \, , \cr
H_{0+}^{(+)} &= {a^+ F_\pi \over k^2 - M_\pi^2} \, , \qquad \cr
H_{0+}^{(-)}&= {a^- F_\pi \over k^2 - M_\pi^2} + {1\over 16 \pi m F_\pi } \cr}
\eqno(8)$$
with $a^{\pm}$ the isopin--even and odd S--wave $\pi N$ scattering lengths.
The form of the pion pole term in $H_{0+}^{(\pm )}$ can easily be understood
from the fact that as $k \to q$  one picks up as a residue the forward $\pi N$
scattering amplitude which at threshold is expressed in terms of the two
S--wave scattering lengths. The relation between axial pion production and
the $\pi N$ scattering amplitude has also been elucidated by Adler in his
seminal work [1]. To leading order in the chiral expansion, i.e. from the
tree graphs calculated with ${\cal L}_{\pi N}^{(1)}$, one finds
$L_{0+}^{(-)} = -1 / ( 8 \pi F_\pi )$ and
$H_{0+}^{(-)} = a^- F_\pi / ( k^2 - M_\pi^2)$, with $a^- = M_\pi / (8 \pi
F_\pi^2 )$ Weinberg's celebrated result for $a^-$ [16].
To this order, all other multipoles vanish.

To work out the corrections at order ${\cal O}(q^3)$, it it mandatory to
perform a complete one--loop calculation with insertions from
${\cal L}_{\pi N}^{(1)}$ and the tree diagrams with exactly one insertion
from ${\cal L}_{\pi N}^{(3)}$. One also has to consider tree graphs with
two insertions from ${\cal L}_{\pi N}^{(2)}$ with a nucleon propagator,
which scales as $1/q$, in between.
The $k^2$--dependence (apart from the pion pole terms in $H^{(\pm)}_{0+}$)
is entirely given by the
few diagrams shown in fig.2. The resulting low--energy theorems for the
various S--wave multipoles are
$$\eqalign{
  M_{0+}^{(+)}
 &= {\sqrt{M_\pi^2 - k^2} \over 16  \pi m F_\pi}
\biggl\lbrace g_A^2 + C_M^{(+)} M_\pi \biggr\rbrace + {\cal O}(q^3 ) \cr
M_{0+}^{(-)}
 &= {\sqrt{M_\pi^2 -k^2} \over 16 \pi m F_\pi} \biggl\lbrace G_M^V(k^2-M_\pi^2)
 - g_A^2 + C_M^{(-)} M_\pi \biggr\rbrace + {\cal O}(q^3) \cr} $$
$$\eqalign{
L_{0+}^{(+)} &= {1 \over 3 \pi M_\pi F_\pi} \biggl\{ \sigma(k^2 - M_\pi^2) -
{1 \over 4} \sigma (0) \biggr\} - {a^+ F_\pi \over M_\pi} - {g_A^2
M_\pi \over 16 \pi m F_\pi} + C^{(+)}_L M_\pi^2 + {\cal O}(q^3) \cr
L_{0+}^{(-)} &= { 1 \over 8 \pi F_\pi} \biggl\lbrace -G_E^V (k^2 - M_\pi^2)
+{M_\pi\over2m}  (g_A^2 + 1) - {k^2 \over 8m^2} \biggr\rbrace + C_L^{(-)}
M_\pi^2 +  {\cal O}(q^3 )  \cr
H_{0+}^{(+)} &= {a^+ F_\pi \over k^2 - M_\pi^2} + {\sigma(0) - \sigma(k^2 -
M_\pi^2 ) \over 12 \pi F_\pi (k^2 - M_\pi^2) } + C_H^{(+)}M_\pi +{\cal O}(q^2)
\cr  H_{0+}^{(-)} &= {a^- F_\pi \over k^2 - M_\pi^2} + {M_\pi [G_E^V (k^2 -
M_\pi^2 ) - 1 ] \over 8 \pi F_\pi (k^2 - M_\pi^2) } +{1 \over 16 \pi m F_\pi} +
C_H^{(-)}M_\pi + {\cal O}(q^2) \cr} \eqno(9)$$
Here, the constants $C^{(\pm )}_{H,L,M}$ subsume numerous $k^2$--independent
kinematical, loop and counterterm corrections (the latter ones stem mainly
from
${\cal L}_{\pi N}^{(3)}$) which we do not need for the following discussion and
which are difficult to pin down exactly. There is, however, one exception to
this. The chiral Ward identity
$$ \partial^\mu A^b_\mu = \hat{m} \, \bar{q} i \tau^b \gamma_5 q \sim M_\pi^2
\eqno(10)$$
demands that $k_0 L_{0+} + k^2 H_{0+} \sim M_\pi^2$ and thus with $k_0 = m(2\mu
+ \mu^2 + \nu)/2(1+\mu)$
$$\eqalign{
C_H^{(+)} &= { a^+ F_\pi \over 2m M_\pi^2} - {\sigma(0) \over 8 \pi
M_\pi^2 m F_\pi} + {g_A^2 \over 32 \pi m^2 F_\pi} = { c_2 + c_3  \over 4 \pi m
F_\pi} \cr C_H^{(-)} &= -{2 g_A^2 + 5 \over 64 \pi  m^2 F_\pi}\,\,. \cr}
\eqno(11)$$
The numerical values of the constants are $C_H^{(+)} = -1.0$ GeV$^{-3}$ and
$C_H^{(-)} = 0.5$ GeV$^{-3}$ (using $M_\pi = 139.57$ MeV, $m = 938.27$
MeV, $F_\pi = 93$ MeV, $g_A = 1.26$, $a^+ = -0.83 \cdot 10^{-2} / M_\pi$
and $\sigma (0) = 45$ MeV [17]).
The argument of the various nucleon form factors in (9) is the threshold value
of the invariant momentum transfer squared
$$ t_{\rm thr} = (q-k)^2_{\rm thr} = {k^2 - M_\pi^2 \over 1+ \mu} = k^2 -
M_\pi^2 + {\cal O}(q^3)\,.    \eqno(12)$$
The $G_{E,M}^V (t)$ are the conventional isovector electric and magnetic
nucleon Sachs form factors in the one loop
approximation. They can easily be constructed from the $F_{1,2}^V (t)$ given in
[13] together with the one loop result for $\sigma(t)$ in heavy
baryon CHPT [13]. The structure of the low--energy theorems for the threshold
mulitpoles reveals corrections which are linear in the pion mass and are
thus non--analytic in the quark masses. This is similar to the effect on the
electric dipole amplitude $E_{0+}$ for pion photoproduction first noticed in
ref.[2]. Of particular interest is the low--energy theorem for $L_{0+}^{(+)}$
where one has the following slope at the photon point $k^2 = 0$
$$ {\partial L_{0+}^{(+)} \over \partial k^2}\bigg|_{k^2 = 0} =
{\sigma'(-M_\pi^2) \over 3 \pi M_\pi F_\pi} + {\cal O}(M_\pi) = {g_A^2 \over
128 \pi^2 F_\pi^3 } \biggl( {6\over 5} - \arctan{1\over 2} \biggr) + {\cal
O}(M_\pi)\,.   \eqno(13)$$
It is very interesting to note that although $L_{0+}^{(+)}$ vanishes
identically in the chiral limit $M_\pi = 0$ the slope at $k^2 = 0$ stays
finite. The formal reason for this behaviour is the
non--analytic dependence of $L_{0+}^{(+)}$ on $M_\pi$ which does not allow to
interchange the order of taking the derivative with respect to $k^2$ at $k^2 =
0$ and the chiral limit. The same phenomenon was observed in [5] where the
electroproduction threshold multipole $E_{0+}^{(-)}$ has been investigated.
Notice also that for $k^2 \simeq 0$ and assuming that $C_L^{(+)}$ of
the order of 1 GeV$^{-3}$, the term proportional to the scalar form factor
$\sigma(-M_\pi^2) - \sigma(0) / 4$
dominates the behaviour of $L_{0+}^{(+)}$ using the numbers from
the recent analysis of Gasser,
Leutwyler and Sainio [17].\footnote{$^{2)}$}{We use a mean square
scalar radius of 1.6
fm$^2$ to estimate $\sigma(-M_\pi^2)$ [17].}
In principle, an accurate measurement of this
particular multipole in weak pion production allows for a new determination of
the elusive nucleon scalar form factor and the $\pi N$ $\sigma$--term.
In contrast to this, the behaviour of $H^{(+)}_{0+}$
is dominated by the pion
pole term proportional to $a^+$. At $k^2 = 0$, one finds $H^{(+)}_{0+} =
32.8 \, {\rm GeV}^{-2} + ( \sigma(0) - \sigma (-M_\pi^2) ) \cdot 14.1
\, {\rm GeV}^{-3} - 0.14 \, {\rm GeV}^{-2}$. The uncertainty in $a^+$,
$\delta a^+ = \pm 0.38 \cdot 10^{-2}/M_\pi$, gives as large a contribution
as the term proportional to the scalar form factor. Finally, we should mention
that
the analysis presented here is not sufficient to estimate the higher order
corrections to the relations (9). This problem has yet to be addressed and can
most probably be resolved by combining dispersion theory with the chiral low
energy constraints.
\bigskip
\noindent 5. We find it furthermore instructive to compare these one loop
results with what one gets from soft pion theory which incorporates PCAC
and current algebra supplemented by the addition of form
factors (see e.g. [8]). This method is frequently used in the literature
in the context of pion photo- and electroproduction theorems. It was
applied by Adler in his extensive study of neutral--current induced
threshold pion production [18].
The pertinent
threshold multipoles are given by the matrix element of the isovector vector
current and the nucleon pole graphs in the limit $q \to 0$,
$$\eqalign{
M_{0+}^{(+)}
 &= {\sqrt{-\nu} \over 16  \pi  F_\pi}{ 4 - \nu \over 4 - 2\nu}
 g_A G_A (k^2)  \cr
M_{0+}^{(-)}
 &= {\sqrt{-\nu} \over 16 \pi F_\pi} \biggl\lbrace G_M^V (k^2) - {4 - \nu \over
4 - 2 \nu } g_A G_A(k^2)  \biggr\rbrace \cr
L_{0+}^{(+)} &= H_{0+}^{(+)} = 0 \cr
L_{0+}^{(-)} &= -{G_E^V (k^2)  \over 4 \pi F_\pi \sqrt{4-\nu }}
 = - 2mH_{0+}^{(-)}\,\,. \cr} \eqno(14)$$
Notice again that the quantities appearing in (14) are given by their values in
the chiral limit. The results (14) are exact for massless pions and agree with
the ones given in (9) since the S--wave $\pi N$ scattering lengths $a^\pm$ as
well as $\sigma(t) / M_\pi$ vanish in the chiral limit (remember that
$\sigma(t) \sim \hat m \sim M_\pi^2$). Clearly the soft pion approach can
not give the most interesting corrections related to the scalar form factor of
the nucleon since this quantity is vanishing in the chiral limit $\hat m = 0$.
The relevant new information contained in (9)  can only be determined by an
explicit loop calculation in CHPT.
\bigskip
\noindent 6. Let us summarize the pertinent results of this investigation. We
have used heavy baryon chiral perturbation theory to investigate threshold
axial pion production. At threshold, there are six multipoles whose chiral
expansion can be expressed in terms of isovector electroweak     form
factors, $\pi N$ S--wave scattering lengths and the nucleon scalar form factor
$\sigma(t)$. In particular, we have found that the $k^2$--dependence of
$L_{0+}^{(+)}$ is due to the scalar form factor $\sigma(k^2 - M_\pi^2)$. This
opens the possibility of another determination of this fundamental
quantity. In the standard model, the axial part of the weak neutral
current is the third component of the isovector axial current. To see
this most interesting correction, one should therefore consider neutral
neutrino  reactions like $\nu p \to \nu p \pi^0$ (in that case the
zero lepton mass approximation is justified).
First, however, a complete calculation involving also the
isovector
vector current\footnote{$^{3)}$}{The chiral expansion of the nucleon isovector
vector form factors has been discussed in [13,14].}
has to be performed to find out how cleanly one can
separate this multipole in the analysis of neutrino--induced single
pion production. For that, it will be mandatory to include the
$\Delta$ resonance since the presently available  data are concentrated
around this mass region [19].
Finally, we point out that Adler's
relation between weak single pion production and the elastic neutrino--nucleon
cross section at low energies [18] is also modified by the novel term
proportional to the scalar form factor of the nucleon. Clearly, when
investigating these questions one has to include the chiral corrections
to the vector current matrix element of single pion production obtained
in ref.[3].
Work along these lines is under way.
\bigskip
 \bigskip
We are grateful to Gerhard Ecker for some useful comments.
\bigskip \bigskip
{\bf REFERENCES} \bigskip
\item{1.}
S.L. Adler,
{\it Ann. Phys. (N.Y.)\/} {\bf 50}
(1968) 189. \smallskip
\item{2.}V. Bernard, J. Gasser, N. Kaiser, and Ulf-G. Mei{\ss}ner,
{\it Phys. Lett.\/} {\bf B268} (1991) 291.
\smallskip
\item{3.}V. Bernard, N. Kaiser, and Ulf-G. Mei{\ss}ner, {\it Phys. Lett.\/}
{\bf B282} (1992) 448.
\smallskip
\item{4.}V. Bernard, N. Kaiser, and Ulf-G. Mei{\ss}ner, {\it Nucl. Phys.\/}
{\bf B383} (1992) 442.
\smallskip
\item{5.}V. Bernard, N. Kaiser, and Ulf-G. Mei{\ss}ner, {\it Phys. Rev.
Lett.\/}
{\bf 69} (1992) 1877.
\smallskip
\item{6.}V. Bernard,  N. Kaiser, T.-S. H Lee and
Ulf-G. Mei{\ss}ner, preprint BUTP-93/23 and CRN 93-45, 1993.
\smallskip
\item{7.}S.L. Adler and Y. Dothan,
{\it Phys. Rev.\/} {\bf 151} (1966) 1267.
\smallskip
\item{8.}Y. Nambu and D. Luri\'e, {\it Phys. Rev.\/} {\bf 125}
(1962) 1429;

Y. Nambu and E. Shrauner, {\it Phys. Rev.\/} {\bf 128}
(1962) 862.
\smallskip
\item{9.}S. Weinberg,
{\it Physica\/} {\bf 96A}
(1979) 327.
\smallskip
\item{10.}J. Gasser and H. Leutwyler, {\it Ann. Phys. (N.Y.)\/}
 {\bf 158} (1984) 142.
\smallskip
\item{11.}Ulf-G. Mei{\ss}ner,
{\it Rep. Prog. Phys.\/} {\bf 56} (1993) 903.
\smallskip
\item{12.}E. Jenkins and A.V. Manohar, {\it Phys. Lett.\/} {\bf B255} (1991)
558.
\smallskip
\item{13.}V. Bernard, N. Kaiser, J. Kambor
and Ulf-G. Mei{\ss}ner, {\it Nucl. Phys.\/} {\bf B388} (1992) 315.
\smallskip
\item{14.}J. Gasser, M.E. Sainio and A. ${\check {\rm S}}$varc, {\it Nucl.
Phys.\/} {\bf B307} (1988) 779.
\smallskip
\item{15.}V. Bernard, N. Kaiser and Ulf-G. Mei{\ss}ner,
{\it Phys. Lett.\/} {\bf B309} (1993) 421.
\smallskip
\item{16.}S. Weinberg, {\it Phys. Rev. Lett.\/} {\bf 17} (1966) 616.
\smallskip
\item{17.}J. Gasser, H. Leutwyler and M.E. Sainio, {\it Phys. Lett.\/}
{\bf B253} (1991) 252,260.
\smallskip
\item{18.}S.L Adler, {\it Phys. Rev. Lett.\/} {\bf 33} (1974) 1511;
{\it Phys. Rev.\/} {\bf D12} (1975) 2644.
\smallskip
\item{19.}S.J. Barish et al., {\it Phys. Rev.\/} {\bf D19}
(1979) 2521;

M. Pohl et al., {\it Lett. Nuovo Cimento}  {\bf 24} (1979) 540;

N.J. Baker et al, {\it Phys. Rev.\/} {\bf D23}
(1981) 2495.
\smallskip
\bigskip \bigskip
{\bf FIGURE CAPTIONS} \bigskip
\item{Fig.1}Basic Feynman diagram for axial pion production.
\medskip
\item{Fig.2}One loop diagrams which lead to the momentum dependence
of eqs.(9). Solid, dashed and wiggly lines denote nucleons, pions and the
isovector axial source ($e.g.$ the $W$-boson), in order.
\smallskip

\vfill \eject \end